# Predicting failure characteristics of structural materials via deep learning based on nondestructive void topology

*Leslie Ching Ow Tiong,[1][†] Gunjick Lee,[2][†] Seok Su Sohn,[2][*] and Donghun Kim[1][*]*

[1]Computational Science Research Center

Korea Institute of Science and Technology, Seoul 02792, Republic of Korea

[2]Department of Materials Science and Engineering

Korea University, Seoul 02841, Republic of Korea

[†]These authors contributed equally.

*Correspondence to: donghun@kist.re.kr (D.K.); sssohn@korea.ac.kr (S.S.S.)

# Abstract

Accurate predictions of the failure progression of structural materials is critical for preventing failure-induced accidents. Despite considerable mechanics modeling-based efforts, accurate prediction remains a challenging task in real-world environments due to unexpected damage factors and defect evolutions. Here, we report a novel method for predicting material failure characteristics that uniquely combines nondestructive X-ray computed tomography (X-CT), persistent homology (PH), and deep multimodal learning (DML). The combined method exploits the microstructural defect state at the time of material examination as an input, and outputs the failure-related properties. Our method is demonstrated to be effective using two types of fracture datasets (tensile and fatigue datasets) with ferritic low alloy steel as a representative structural material. The method achieves a mean absolute error (MAE) of 0.09 in predicting the local strain with the tensile dataset and an MAE of 0.14 in predicting the fracture progress with the fatigue dataset. These high accuracies are mainly due to PH processing of the X-CT images, which transforms complex and noisy three-dimensional X-CT images into compact two-dimensional persistence diagrams that preserve key topological features such as the internal void size, density, and distribution. The combined PH and DML processing of 3D X-CT data is our unique approach enabling reliable failure predictions at the time of material examination based on void topology progressions, and the method can be extended to various nondestructive failure tests for practical use.

# Introduction

Structural materials are designed to withstand the various damage factors that exist in application environments. Despite considerable research on material fatigue damage in recent decades, accurate predictions of the material lifetime remain challenging due to the complexity and high variability of real-world environments[1–8]. In fact, unexpected accidents caused by material failure have recently occurred, including aircraft disintegration in the air, building/bridge collapses, gas pipeline explosions, and container ships sinking[9–13]. These accidents have resulted in significant loss of life and property and, in severe cases, hundreds of casualties. These losses can be minimized by predicting when a material will fail and adopting preventative actions before material failure. Thus, methods for quantitatively predicting failure-related properties in environments where various factors, including load, heat, and corrosion, may comprehensively damage materials are crucial[14–18].

Models for predicting the failure behaviors and lifetimes of structural materials have been extensively studied for decades. Examples primarily include the mechanics-based finite element method (FEM) combined with the Gurson-Tvergaard-Needleman model, the Gunawardena model, and the continuum damage mechanics model[19–25]. The FEM has facilitated fatigue life predictions based on inputs such as the variation in yield locus under repeated loading cycles and the corresponding stress-life cycle curves ($S$-$N$ curves). In addition, the geometry of the material can be considered, and the FEM has no constraints on the loading and boundary conditions applied to the materials[19,26–29]. However, this mechanics-based approach is limited in its ability to represent microstructural defects that develop differently depending on the exposed environmental factors. In this regard, a crystal plasticity FEM (CPFEM) has recently been explored to investigate the influences of microscopic scale features, including the evolution of microstructural defects such as voids and grains[28–32]. This method elucidates the stress–strain states near defects such as inclusions and pores, as well as their effects on crack propagation and fatigue lifetime[33].

Nevertheless, it remains difficult to reflect the statistical information of defects, such as the evolution of their fraction, size, and distribution at each failure progression due to prohibitive computational costs, limiting the connections between microstructures and macro-properties at different length scales (e.g., microstructural defect—strain state—fatigue lifetime). In this respect, data-driven machine learning (ML) approaches can overcome this scale gap since ML is capable of understanding correlations within datasets regardless of their length or time scale. Owing to these benefits, various ML techniques, such as random forests, support vector machines, and neural networks, have recently been used to predict the mechanical properties of metal systems, including their hardness[34,35], tensile properties[36,37], and crack propagation[38,39]. However, these ML models rely solely on technical parameters such as experimental processing conditions as inputs, and no ML models have yet to be developed based on microscopic defect states that are obtained nondestructively at each failure progression, which is both a fundamental cause of failure and a critical requirement for predicting failure characteristics in real-world application environments.

In this work, we propose a method that uniquely combines X-ray measurements, persistent homology (PH), and deep learning (DL) to predict the failure characteristics of structural materials based on their defect states at the time of material examination. In this study, X-ray computed tomography (X-CT) scanning, a nondestructive scanning method, is used to monitor the evolution of the defect states. Our proposed method takes X-CT images as its only input and outputs the failure-related properties, such as the local strain and fracture progress. The effectiveness of the method is demonstrated using two fracture datasets, which were produced by tensile and fatigue testing of ferritic low alloy steel as a representative structural alloy. When trained/tested on 135 tensile testing data points, the method predicted local strains with a mean absolute error (MAE) of 0.09. Similarly, when applied to 100 fatigue testing data points, the method predicted the fracture progress with an MAE of 0.14. These values are significantly lower than the MAEs of 0.55 and 0.57 obtained for cases without PH results. The remarkable enhancements are mainly attributed to the PH process, which transforms irregular and noisy three-dimensional (3D) X-CT images into two-dimensional (2D) persistence diagrams (PDs) that concisely describe the topological features of internal voids, such as their size, density, and distribution. The developed workflow could be easily extended to other nondestructive scanning methods, including phased array ultrasonic testing, for practical use.

# Results

**Model description.** Fig. 1 illustrates the overall scheme of our deep learning method. This method takes X-CT images of structural materials at the time of examination as its only input and outputs the failure-related properties. For model development and validation, in this study, we construct two types of fracture datasets: one produced by tensile mechanical testing and one produced by fatigue mechanical testing. The two datasets output different predictions: the tensile test outputs the local strain, and the fatigue test outputs the fracture progress. X-CT scanning can identify the positions and shapes of vacant regions, such as voids or cracks, in structural materials. Next, the X-CT images are processed by a combination of persistent homology and deep learning. The 3D X-CT images are transformed into 2D PDs via PH processing. This PH introduction step is an important innovation of this work since the patterns in 3D X-CT images are often too irregular and noisy to be directly used as the input of the following machine learning process. On the other hand, 2D PDs compressively capture and quantify key topological and geometrical features of defects in materials, such as their size, shape, density, and distribution. Thus, instead of the raw X-CT data, the transformed PDs are used as inputs to the ML process to predict failure-related properties, such as local strains or the fracture progress.

In Fig. 1, among all ML models tested in this work, the deep multimodal learning (DML) model is representatively schematized due to its outstanding performance over benchmark models. Unlike the other models, which rely on a single input source, the DML model is multimodal and accepts two types of inputs: the PD images themselves and PD-extracted metrics. The PD-extracted metrics include the metrics extracted from PDs that quantify the void density, intervoid distance, and heterogeneity of void distributions. For the DML model to process the heterogeneous inputs of image data and value-type data, convolutional neural networks (CNNs) and deep neural networks (DNNs) are used in parallel. We found that a DML model with two input sources considerably outperformed other benchmark models with a single input source. The PH and DML processing of the 3D X-CT data is our unique approach and key strategy for quantitatively predicting failure-related properties at the time of material examination. The detailed processes for the PD computations and ML model development are described in Figs. 2 and 5, respectively, along with related explanations.

**PH process for quantifying void topology.** Topological data analysis (TDA) is an emerging mathematical tool that uses algebraic topology principles to comprehensively measure shapes in datasets. TDA has shown promise in various application domains, including materials science, as it detects essential topological features embedded in a system by tracking the lifetime of *holes* in each dimension, such as connected components, loops, and voids[40–44]. TDA is often narrowly used to describe a particular subfield, namely, PH. In this work, PH is used to identify and quantify topological and geometric features in 3D X-CT images[45].

We compute the $0^{th}$ PD via the PH process, and the intuitive idea underlying the PH process is illustrated in Fig. 2a. In a schematic example where the 3D X-CT scanning data contain multiple voids, the data can be mathematically regarded as 3D binary format data with void regions *versus* the remaining background region. The first step in the PH process is *pixelation,* where each pixel in the 3D space is assigned a number, such as $t_0$, $t_1$, $t_2$, $t_3$, $t_4$ and so forth (with pixels inside voids assigned values of $\leq t_2$ and pixels outside voids assigned values of $>t_2$) to be used in the subsequent PD computations. The rule for assigning these numbers is known as the Manhattan distance, and the number assigned at boundary surfaces ($t_2$ in Fig. 2a) is usually referenced to zero[41,46]. The next step is *filtration,* in which the union of pixels with assigned numbers less than a threshold value ***T*** is analyzed. When the threshold ***T*** is increased from $t_0$ to $t_4$, some islands appear (birth) and disappear (death), and the event times of these birth and death pairs are encoded in the $0^{th}$ PD. When ***T***=$t_3$ or $t_4$, the neighboring islands are observed to be in contact and combined into one body, and the smaller island is treated as a death. For the example shown in Fig. 2a, the (birth, death) pairs of ($t_0$, $t_3$) and ($t_1$, $t_4$) are encoded in the $0^{th}$ PD.

Fig. 2b illustrates the information stored in the $0^{th}$ PDs using 15 exemplary cases with different void topologies. Instead of PD images, we introduce birth (***b***) and lifetime (***d-b***) histograms, which can be obtained from simple reconstructions of the PD results, to more intuitively understand how different void topologies are reflected in the PD results. If the number of voids in a sample is assumed to be constant, the void topology can be varied by modifying three features: the void size, intervoid distance, and void distributional heterogeneity. The variations in each feature are schematized in three rows in Fig. 2b. For the samples in the top row, the void topology differs only in terms of the void size, with the relative positions and distributions of the voids unchanged. Since larger voids appear earlier in the *filtration* process, peaks in the birth (***b***) histograms are shifted to the left (smaller birth values) for cases with larger voids. Next, we compare samples in the middle row of Fig. 2b, which exhibit different distributional features, namely, different void-pair distances. Since distant void pairs have longer void lifetimes in the *filtration* process, these differences should be reflected in the lifetime (*d-b*) histograms, and peaks in the histograms are shifted to the right (larger lifetime values) for distant void pairs. The last feature is the void distributional heterogeneity, which is schematized in the bottom row of Fig. 2b. The difference in heterogeneity is related to the degree of variation in the lifetime (*d-b*) histograms. Void lifetime values should be similar for homogeneous distributions, whereas these values vary substantially for heterogeneous cases. Overall, the PH processing of the 3D X-CT data extracted and quantified key topological features, including the void density, void size and some distributional features such as the intervoid distance and void heterogeneity. These PH capabilities are evaluated using two experimental fracture datasets obtained from different mechanical tests: one obtained from tensile tests and the other obtained from fatigue tests of metallic materials.

**Dataset generation from tensile tests.** We generated the fracture datasets used to develop the method shown in Fig. 1. Ferritic low alloy steel, a representative structural material, is selected as the test material in this work[47,48]. The specimens used for the tensile test are plate-shaped, with a gauge length of a 6.4 mm and a thickness of 1.5 mm (inset of Fig. 3a). Tensile tests with a strain rate of $10^{-3}$ s$^{-1}$ were performed on 15 ferritic steel specimens at room temperature. The total elongation when the sample broke was 40.7±3.7% for these 15 samples, as shown in Fig. 3a and Supplementary Fig. 1. The digital image correlation (DIC) technique was used to measure the local strains of the samples, as shown in the local strain map in Fig. 3b. The central part of the specimen near the fracture point was highly strained, exhibiting local strains of 100–130%. As the local strain varies substantially at different positions of the specimen, X-CT scanning was performed on multiple regions of the specimens (6–8 parts divided along the 6 mm length near the fractured surface).

A total of 135 data points were collected in the tensile dataset, and each data point consists of X-CT images and the corresponding local strain value of the X-CT-scanned part of the specimen. Fig. 3c shows an overview of the tensile dataset, displaying the local strain of each data point as a function of the void density (# mm$^{-3}$) and average intervoid distance (***D***). The void density is defined as the number of voids divided by the scan volume, and $D$ is defined in the Methods section as the geometric mean of the lifetime ($d$-$b$) values in the PDs. The whole dataset can be roughly classified into regions with low, mid, and high local strains. As the strain increases from low to high, the void density tends to decrease, whereas $D$ tends to increase. This tendency is visually confirmed in the internal void views revealed by X-CT scanning, where the voids decrease in number and increase in size, resulting in a larger value of $D$ for high strain cases. The observed evolution of the void topology can be adequately explained by the void coalescence phenomenon, which is known to occur in ductile alloys under accumulated plastic deformations[49,50].

While Fig. 3c displays the overall dataset, Fig. 4 focuses on a specific specimen as an example for a more detailed PH analysis. A fractured ferritic steel sample is divided into four parts for the X-CT measurements (parts 1–4), with part 4 located nearest to the fractured surface. Each part is represented by different local strain values, namely, 0.17 (part 1), 0.23 (part 2), 0.40 (part 3), and 0.82 (part 4). In Fig. 4b, the $0^{th}$ PDs for parts 1—4 are computed via the PH process introduced in Fig. 2a. These PDs are noticeably different from one another. In particular, as the local strain increases, the dot colors inside the red circles in Fig. 4b gradually change from red to purple, indicating that the density of the small voids (≤3 μm) substantially decreases.

In Figs. 4c and 4d, to more intuitively understand the quantitative difference in the void topology, birth ($b$) histograms and lifetime ($d$-$b$) histograms are built based on the PDs shown in Fig. 4b. Since the birth values are directly related to the void size (in this case, size ≈ -3×$b$ μm because the unit length for each pixel in the PH process is 3 μm), the birth histogram summarizes the void size and density statistics. As the local strain increases (part 1 → part 4), the density of smaller voids (≤3 μm) rapidly decreases (8,638 → 6,047 → 2,634 → 304 mm$^{-3}$), whereas that of larger voids (3–21 μm) increases. This trend is consistent with the void coalescence phenomenon, in which small voids are combined into larger voids under heavy loads. Next, as the lifetime value represents the distance between neighboring voids (in this case, intervoid distance ≈ 3×($d$-$b$) μm), the lifetime histogram presents the statistics of the void distributional features. $D$, which is defined as the geometric average intervoid distance, increases as the local strain increases ($D$ = 16.9 → 18.0 → 23.0 → 32.8 μm). Additionally, we introduce the void heterogeneity ($V$), which is defined as the degree of variation in the lifetime histograms. The equations for $D$ and $V$ are provided in the Methods section. $V$ also increases ($V$ = 3.8 → 4.7 → 5.6 → 10.7 μm) with increasing local strain.

This PH analysis of the void topology of a ferrite specimen reveals that our method can quantify key void information such as size, density, and distributional features of intervoid distance and heterogeneity; thus, PD images and PD-extracted metrics ($D$ and $V$) are suitable inputs for the subsequent ML studies.

**Machine learning tensile dataset.** We performed ML experiments with the tensile dataset to classify or predict local strains based on the PH results. Five ML models (Models **I–V**) are schematized in Fig. 5a. Although these models all output the local strain of the tested sample, the ML algorithms, architectures, and inputs all differ. The ML algorithms include multivariable linear regression[51] (MLR), DNN[52], and CNN[53]. In terms of input types, Model **II** differs from the other four models: Model **II** uses the raw X-CT images as inputs, whereas Models **I**, **III**, **IV**, and **V** use the PH results (PD images or PD-extracted metrics) as inputs. Here, the PD-extracted metrics include the following five metrics: void density (for void size ≤3 μm), void density (for void size of 3–6 μm), void density (for void size >6 μm), $D$, and $V$. The void density is categorized into three classes based on the results of Fig. 4c, where the tendency of void density of small (≤3 μm) and larger (3–6 μm or >6 μm) voids differed with local strain variations. For the MLR or DNN cases (Models **I** and **III**), PD-extracted metrics are used as inputs, while the CNN model (Model **IV**) uses the PD images as inputs since the CNN is a specialized algorithm for processing images. For the DML model (Model **V**), Models **III** and **IV** are combined in parallel to utilize both PD images and PD-extracted metrics as inputs. The details of the network configurations are provided in Supplementary Tables 1 and 2.

First, ML classification tasks are performed using the tensile dataset. The local strain values range between 0.10–1.30 and can be categorized into five classes with 0.15 intervals: class #1: 0.10—0.25, class #2: 0.25—0.40, class #3: 0.40—0.55, class #4: 0.55—0.70, and class #5: ≥0.70. The accuracy results of the top-$k$ classifications for Models **I–V** are summarized in Fig. 5b. The top-$k$ accuracy refers to the percentage of cases in which the correct class label appears among the top-$k$ probabilities. Model **V** (DML model) showed an excellent performance, with a top-1 classification accuracy of 84.6% and 100% top-$k$ ($k≥2$) accuracies. Model **IV** (CNN model based on PD images) also achieved a high top-1 accuracy of 80.8%. On the other hand, Models **I** and **III**, which are based on PD-extracted metrics, considerably underperformed, exhibiting accuracies of less than 75%. Model **II**, which is based on raw X-CT images, was inefficient, with a top-1 classification accuracy of only 31%. The very low performance of Model **II**, which does not use any PH-based inputs, demonstrates the effectiveness and necessity of PH processing of noisy X-CT images.

Next, for the prediction tasks, all models are trained to predict, not classify, local strain values. The MAE is used to evaluate the prediction performance. Similar to the classification results, as shown in Figs. 5c-g, Model **V** (DML model) achieved the best MAE of 0.09. Model **III** achieved the second-lowest MAE of 0.12, and Models **I** and **IV** achieved MAEs of 0.42 and 0.33, respectively. Model **II**, which used raw X-CT images as inputs, performs the worst, exhibiting an MAE of 0.55, similar to the classification results, indicating that PH processing is a necessary step to achieve high prediction performance.

Since Model **V** (our DML model) performs the best in both the classification and prediction tasks, it is worth investigating the reasons for these improvements. Model **V** includes two networks to exploit the different features of PD images and PD-extracted metrics. The two modal networks are based on a CNN and DNN; the former focuses on learning the features in the PD image, while the latter is trained on the PD-extracted metrics. The DML model simultaneously accepts and processes these different and complementary inputs and generates a joint feature representation to strengthen the feature activations of the network. This multimodal architecture is more effective at learning complex features than models that rely on a single input source.

**Application to fatigue dataset**. Figs. 3-5 show that the developed method is effective at classifying and predicting local strains using the tensile dataset. Our method was also applied to another fracture dataset, namely, the fatigue dataset, to predict fracture progression. In real application environments, failure-induced accidents occur unexpectedly through fatigue mechanisms; thus, an expansion to a fatigue-driven fracture dataset is necessary for the method to be used as a practical and universal tool. The ferritic low alloy steel was also used in the fatigue testing, and the specimens were plate-shaped, with a 3 mm gauge width and a 1.5 mm thickness. The detailed dimensions of the fatigue testing specimens are illustrated in Supplementary Fig 2. To induce void-driven fatigue fractures, the original ferrite specimens were intentionally prestrained with 10% tensile deformations before fatigue testing, causing small homogeneous voids to form within the specimens[54]. A fatigue uniaxial load was applied with a maximum stress of 700 MPa, an *R*-ratio of 0.1, and a frequency of 10 Hz. Fig. 6a illustrates the process of collecting the fatigue datasets. The fatigue tests were interrupted at cycles of $10^n$ (n = 3.0, 3.5, 4.0, 4.5 and so forth) until the failure point for the X-CT measurements. The fracture progress of each specimen was then determined by dividing the interrupted cycle number (*N*) by the fatigue life cycle ($N_f$) of that specimen. For the exemplary case shown in Fig. 6a, the specimen was interrupted four times before it failed, at progresses of 9%, 28%, 88%, and 97%, and the defect topology evolutions in the gauge region of the specimens were obtained by X-CT scanning.

The fatigue dataset includes a total of 100 data points, and each data point is composed of X-CT images and the corresponding fracture progress. Fig. 6b illustrates the whole fatigue dataset in terms of the fracture progress ($N/N_f$), void density, and **D**. The data distribution can be classified into three categories, with low, mid, and high fracture progress. As the fracture progress increases from low to high, the void density tends to increase, resulting in a decrease in **D**. This tendency can be visually confirmed with the internal void views shown in Fig. 6c, and this tendency is a discriminative feature of ML studies, likely leading to a high prediction accuracy. Figs. 6c-6e illustrate the $0^{th}$ PD images and corresponding birth and lifetime histograms for representative samples with low, mid, and high fracture progress. These PD images differ mainly in the birth value range of -6 to -2, which indicates the appearance of larger voids (3–15 μm) in samples with higher progress. Quantitatively, in the birth histograms (Fig. 6d), as the progress increases, the density of small voids (≤3 μm) increases (877 → 3,311 → 4,826 mm$^{-3}$), while larger voids (3–15 μm) with densities of more than a few tens mm$^{-3}$ typically form when the progress exceeds 90%. In terms of the distributional features stored in lifetime histograms (Fig. 6e), as the progress increases, **D** and **V** both decrease, with values of 40.1 → 27.3 → 18.3 μm and 8.6 → 6.6 → 6.0 μm, respectively. These results are consistent because an increase in the void number should cause the voids to be distributed in a closer and more homogeneous manner. These various PD features, including **D** and **V**, can discriminate samples with different fracture progresses and can thus be used as inputs in ML studies.

ML experiments were carried out on the fatigue datasets to classify or predict the fracture progress. Models **I–V**, which are shown in Fig. 5, were also tested using the fatigue dataset. First, ML classification tasks were performed. The progress values can be categorized into the following four classes: class #1: ≤0.2, class #2: 0.2–0.5, class #3: 0.5–0.9, and class #4: 0.9–1.0. Note that the intervals between the four classes are not identical; instead, the intervals were adjusted to ensure that the amount of data in each category is similar. A uniform data distribution over all classes is necessary for unbiased ML predictions, particularly when the amount of data is small. The top-$k$ classification accuracy results for Models **I**–**V** are summarized in Fig. 6f. Similar to the tensile case, Model **V** (DML model) performs the best, with top-1 and top-2 classification accuracies of 82.8% and 95.7%, respectively. Model **IV** (CNN model based on PD images) achieved the second-best performance, with for top-1 and top-2 accuracies of 79.3% and 92.4%, respectively. Models **I** and **III**, which are based on only PD-extracted metrics (no PD images), underperform, achieving top-1 classification accuracies of less than 70%. Model **II**, which utilizes raw X-CT images as its sole input source (no PH-based inputs), exhibits the worst top-1 accuracy of only 33%, indicating that ML training based on only raw X-CT data is not successful. Comparisons between Model **II** and the other models show that the PH process is an effective approach in machine learning for handling complex and noisy X-CT image data.

Next, for the prediction tasks, Models **I**–**V** are used to predict the fracture progress. Figs. 6g-6k show that, similar to the classification results, Model **V** (DML model) achieved the best MAE of 0.14, and Model **IV** achieved the second-best MAE of 0.18. Models **I** and **III**, which use PD-extracted metrics as inputs, showed worse MAEs of 0.35 and 0.25, respectively, indicating that PD-extracted metrics alone cannot sufficiently capture the essential void-related information required for ML training. Unsurprisingly, Model **II**, which does not use any PH-based inputs, exhibited the poorest MAE of 0.57, again supporting the difficulty of training with raw X-CT images. The performance enhancement of our DML model (Model **V**) over other benchmark models appears to be universal in both the tensile and fatigue datasets. Unlike most other ML models, which rely on single input sources, our DML model benefits from exploiting discriminatory features from two input sources, namely, PD images and PD-extracted metrics.

## Discussion

We confirmed in Results that the combined PH and DL models provides the high prediction performances for failure progresses and embodies the associated failure characteristics. Yet, it was not sufficiently understood which factors regarding the void topology play a prior role for the reliable property predictions. In this regard, in Fig. 7, an occlusion sensitivity analysis (OSA) was performed for Models **III** and **IV** and for the tensile and fatigue datasets. OSA is used to quantify the importance of each input feature and can thereby identify key components learned by the models. Here, an occlusion process refers to partially hiding a specific feature of the input data, and the change in classification accuracy due to the occlusion is evaluated. This analysis reveals the sensitivity of output predictions to occluded input data.

The OSA results for Model **IV** (CNN based on PD images) and Model **III** (DNN based on PD-extracted metrics) are illustrated in Figs. 7a and 7b, respectively. For Model **IV**, the occlusion hides a specific location in the PD images, and the accuracy change due to each occlusion part was computed. Regions corresponding to the cluster of small voids (marked in black) were identified as highly sensitive spots, with occlusions causing an accuracy degradation of 15-35% for both the tensile and fatigue datasets. These results are consistent with previous observations that the void density of small voids (≤3 μm) is substantially affected by local strains (tensile) and fracture progress (fatigue), indicating that the development of small voids is a key learning feature. The OSA result for Model **III** in Fig. 7b reveals additional information. Unlike the CNN case, the occlusion in the DNN model involves changing a specific input data feature to a value of zero, and the effects of these occlusions on the output predictions were quantified. This test revealed that the density of small voids (≤3 um) and $D$ are both critical learning features. The priority of these two features (void density for voids ≤3 μm and $D$) is reversed depending on the dataset used: the void density (≤3 um) is more important for the tensile dataset, while $D$ is more important for the fatigue dataset.

Despite the excellent prediction performance of our DML model, we would like to discuss some strategies for addressing the limitations of our dataset and models. Our datasets were designed in fully controlled environments, with uniaxial tensile and fatigue tests conducted at room temperature. Thus, the developed tool may perform poorly in uncontrolled situations due to large differences with the training data. To overcome this limitation, datasets need to be constructed in more diverse experimental environments, in which the strain rates, fatigue modes, stress amplitudes, $R$-ratios, and temperatures can be crucial parameters. Additionally, in the future, our model can be improved further by introducing an attention mechanism into the DML network.

Standard CNN or DNN models extract features directly from the given input; however, for models with an attention mechanism, the crucial part of the input (as opposed to the entire input) is highlighted, enhancing the inference and prediction performance.

In addition to X-rays, our method can be applied to other types of nondestructive scanning tests, such as ultrasonic imaging measurements. Although we selected X-ray scans to obtain high-resolution characterizations of defects, X-ray scanning often investigates localized areas due to limited permeability issues in highly dense metal systems such as steels. On the other hand, ultrasonic testing offers the benefits of scanning wider and deeper areas and is thus considered a more suitable choice in industrial situations, despite its comparatively lower resolution. In fact, ultrasonic-based scanning, such as phased array ultrasonic testing (PAUT), is the most commonly used nondestructive scanning method in industry. Thus, our method should be expanded to fracture datasets obtained by PAUT for practical use.

## Methods

**Material fabrication.** The ferritic low alloy steel used in this work is composed as follows: (<0.05)C– (<1.7)Mn– (<0.3)Si– (<0.4)(Cr+Mo)– (<0.15)(Ti+Nb+V) (wt.%). The steel plate was homogenized at 1200–1250 °C for 1 h and hot-rolled to 800–850 °C, the finisher dispatch temperature. After that, the plate was cooled and coiled at 500–550 °C. The final thickness of the fabricated plate was 6.5 mm.

**X-CT measurement and image processing.** An X-ray microscope (ZEISS Xradia 520 Versa) was used. The output voltage, power, and step size of the X-CT scanning measurement were 160 kV, 10 W, and 3 μm, respectively. To obtain 3D binary format images (voids *versus* remaining background) from the X-CT data, we used an image processing approach, namely, window leveling, as illustrated in Supplementary Fig. 3. This approach aims to control the contrast within the possible range to remove the background and discern the void parts. As a result, the processed X-CT images are converted to 3D binary format images $\boldsymbol{B}$ as follows:

$$\boldsymbol{B}(u,v) = \begin{cases} 255, & if\ I(u,v) > \alpha \\ 0, & otherwise \end{cases}$$

where $I(u, v)$ represents the pixel intensity, $u$ and $v$ are the pixel coordinates, and $α$ is a threshold that is defined as 128 in this study.

**PD computations.** We used the PH processing API–HomCloud[55] to compute the $0^{th}$ PD for the 3D binary format images. To compute the PD-extracted metrics, we first define $D$, the average intervoid distance, as the geometric mean of the lifetime ($d$-$b$) values, which can be calculated as follows:

$$D = \sqrt[M]{\prod_{i=1}^{M} d_i - b_i},$$

where $M$ is the total number of void pairs and $b_i$ and $d_i$ refer to the birth and death times of the $i^{th}$ void pair. We also define $V$, the heterogeneity, as the degree of lifetime variations, which can be calculated as follows.

$$V = \sqrt[M]{\prod_{i=1}^{M} |(d_i - b_i) - D|}$$

**Machine learning**. For the experimental protocols, the dataset was randomly divided into a training set and a testing set with a ratio of 80:20 with no overlap. The samples in the training set were further divided for cross-validation purposes. The cross-validation protocol was designed as follows: (1) randomly shuffle the training set; (2) divide the training into five groups; (3) use one group as the validation set and the remaining groups as the training set; (4) repeat step 3 every 50 epochs and summarize the model evaluation scores. For the testing scheme, the testing set was used to evaluate the performance of our models. All network layer configurations are summarized in Supplementary Tables 1 and 2.

For the DML model (Model **V**), the networks were trained simultaneously. During training, we used an Adam optimizer[56] with a learning rate of $1.0 \times 10^{-4}$; the weight decay and momentum of the optimizer were defined as $1.0 \times 10^{-8}$ and 0.9, respectively. We defined the total loss[57,58] ($\ell_{total}$) function as the sum of the cross-entropies of the logit vectors, as well as their respective encoded labels, as follows:

$$l_{total} = l(FC_1) + l(FC_2),$$

$$l(FC_*) = -\sum_k^K \sum_c^C L_{kc} \log[\delta(FC_*)_{kc}],$$

$$\delta(FC_*)_{kc} = \frac{\exp^{(FC_*)_{kc}}}{\sum_c^C \exp^{(FC_*)_{kc}}},$$

where * denotes the modal inputs (1 refers to PD images and 2 refers to PD-extracted metrics), $FC$ is a fully connected layer, $L$ denotes the class labels, $K$ is the number of training samples, $C$ is the number of classes, and $\delta$ is the output layer, which is implemented with the softmax[59] function. For the prediction task, we used the following $\ell_{total}$ function:

$$l_{total} = l(\hat{y}_1) + l(\hat{y}_2)$$

$$l(\hat{y}_*) = \sum_{i=1}^n (y_{i,*} - \hat{y}_{i,*})^2$$

where * denotes the modal inputs (1 refers to PD images and 2 refers to PD-extracted metrics), $\hat{y}$ is the predicted value, $y$ denotes the actual value, and $\ell(\cdot)$ is defined as the mean square error loss function.

## Data availability

The data that support the findings of this study are available from the corresponding author upon reasonable request.

## Code availability

The demonstration of the DML model is written in Python and TensorFlow toolkit, which is available at https://github.com/tiongleslie/material-failure-prediction.

# Acknowledgments

This work was supported by Samsung Research Funding & Incubation Center of Samsung Electronics under Project Number SRFC-MA1902-04.

# Author contributions

D.K. and S.S.S. designed and supervised the research. G.L. and S.S.S. prepared experimental materials, and performed fracture tests. G.L. and L.C.O.T collected X-CT images. L.C.O.T. and D.K. performed image processing of X-CT images, PH analysis, and ML training and tests. All authors analyzed the data and contributed to writing the manuscript.

## Competing Interest

We declare that none of the authors have competing financial or non-financial interests as defined by Nature Research.

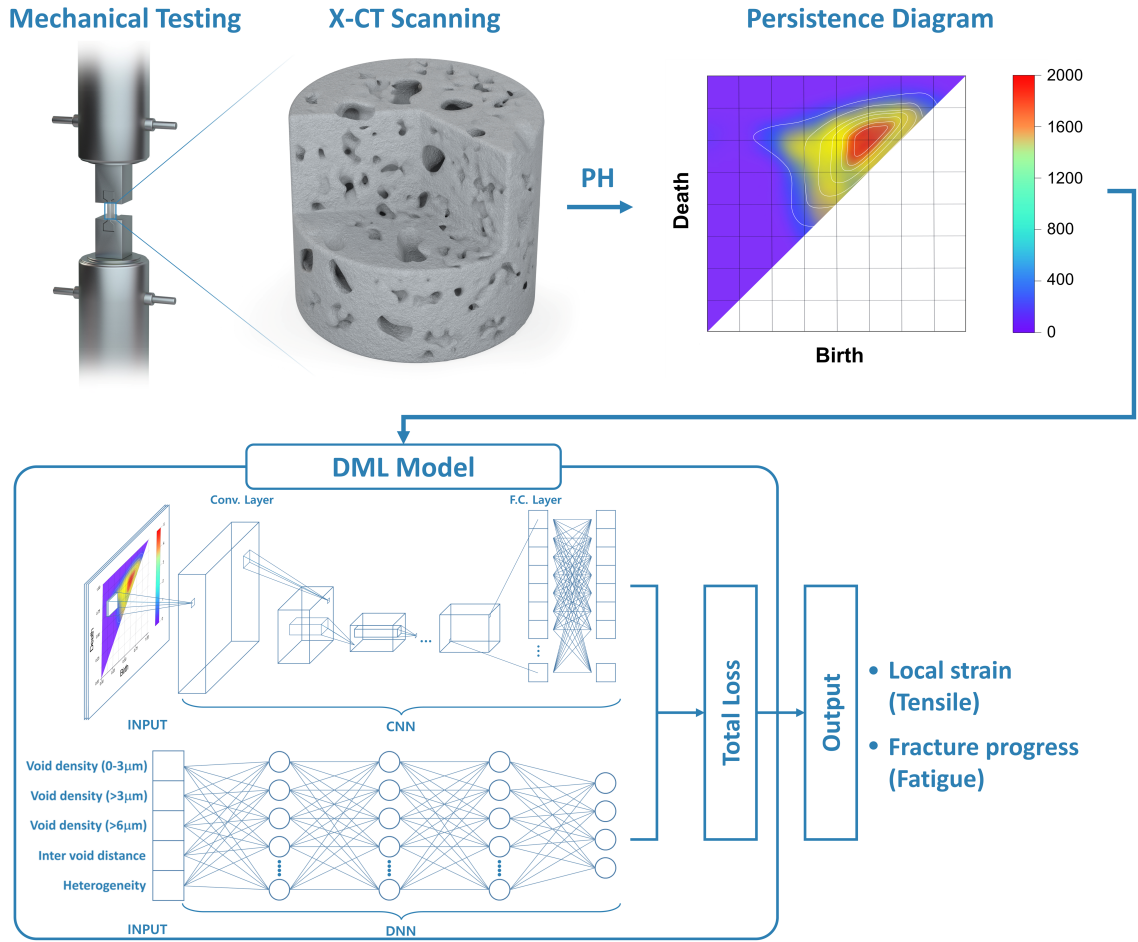

**Fig. 1. Scheme of the proposed method.** The method accepts X-CT images containing defect information as its sole input and outputs the failure-related properties at the time of material examination, such as the local strain for the tensile test and the fracture progress for the fatigue test.

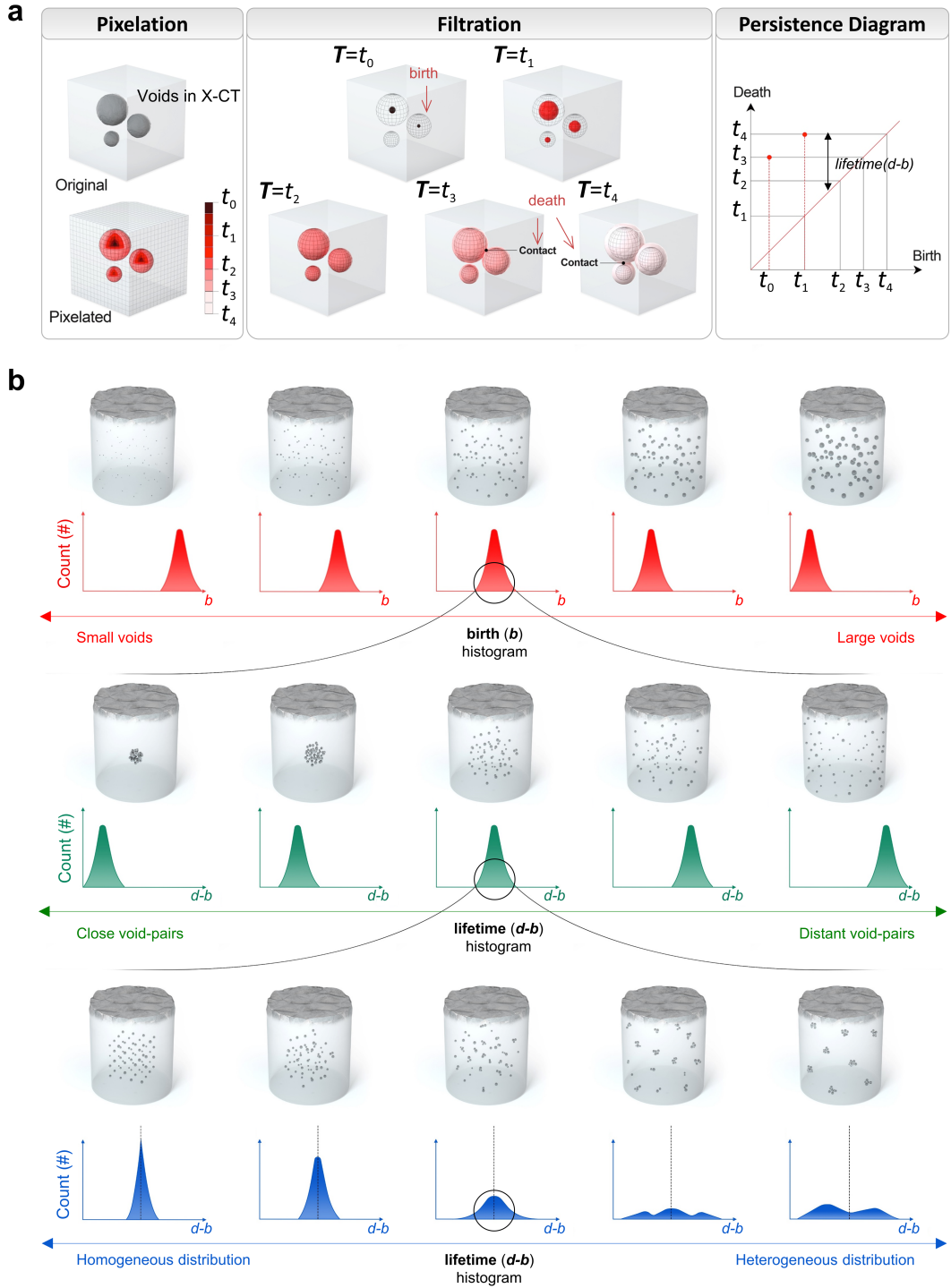

**Fig. 2. PH process for measuring void topological features. a** Scheme describing the PH process and the $0^{th}$ PD computation. **b** Fifteen exemplary cases with different void topologies. The samples in the top row have different void sizes, the samples in the middle row have different intervoid distances, and the samples in the bottom row have different distributional heterogeneities. The corresponding birth (***b***) and lifetime (***d-b***) histograms are displayed.

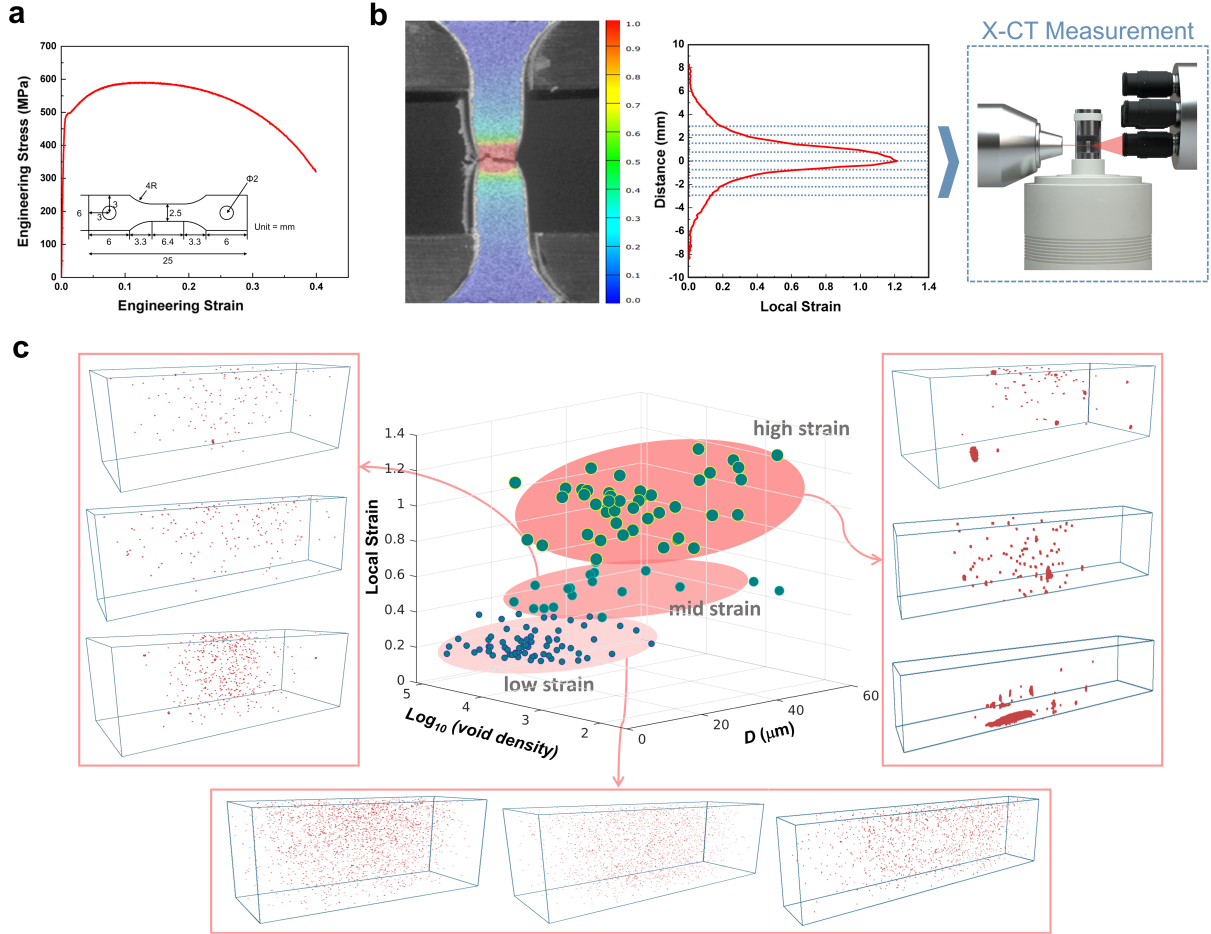

**Fig. 3. A fracture dataset generated from tensile mechanical testing. a** A representative engineering stress–strain curve of a test ferritic steel sample. The specimen dimensions are shown in the inset. **b** Local strain map and X-CT measurement process. An exemplary local strain map of the fractured sample obtained by the DIC technique is shown, with the color bar representing local strain values, as well as the corresponding strain profile. **c** The tensile dataset at a glance. The graph displays the local strain (*z*-axis) as a function of the void density (# mm$^{-3}$, log scale) and ***D*** (μm).

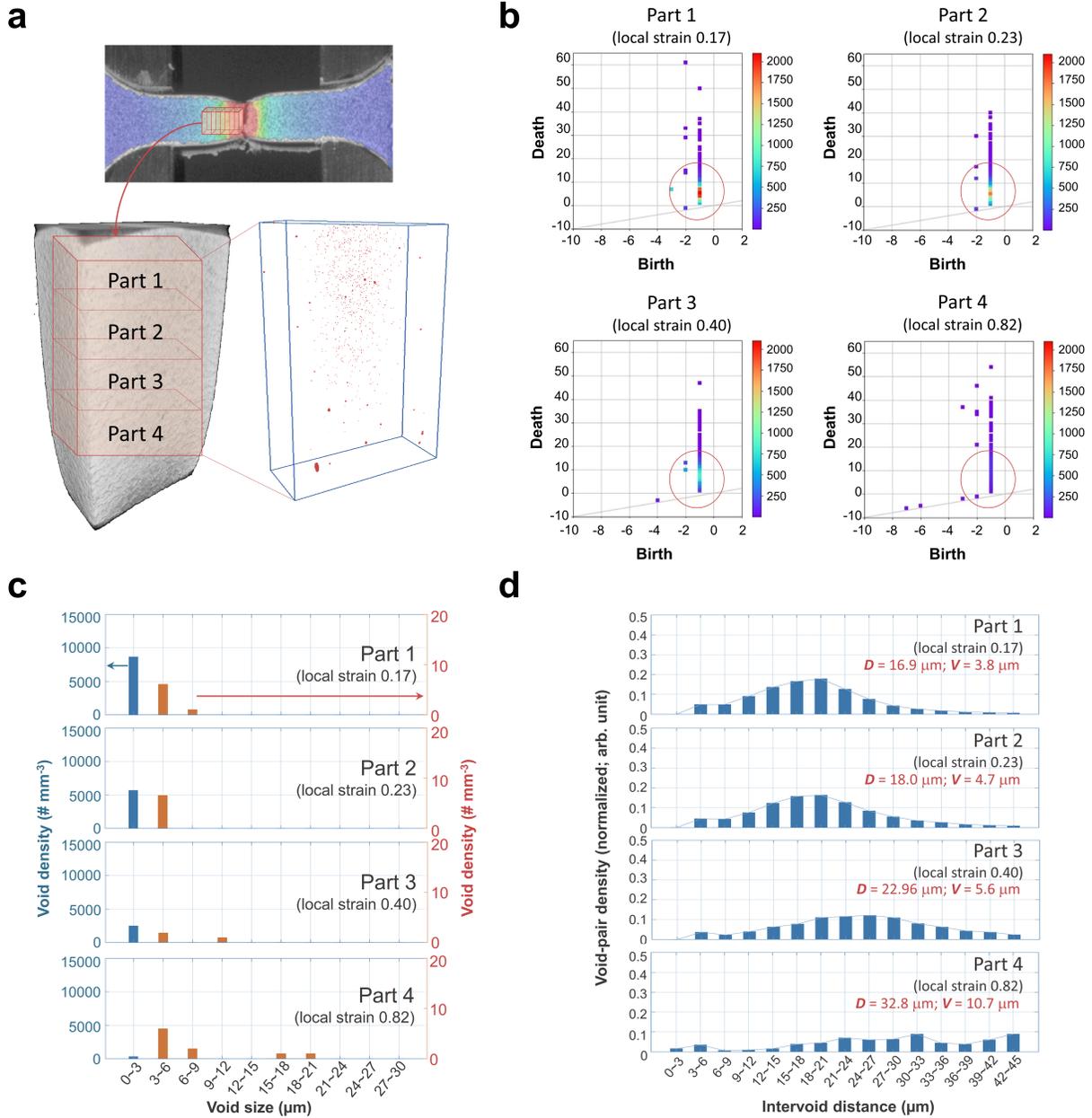

**Fig. 4. PH analysis for a ferrite specimen fractured during tensile mechanical testing. a** 3D structural view of fractured ferrite samples. The DIC result, X-CT overview, and internal void view are shown. **b** The 0$^{th}$ PDs of parts 1–4. **c** Birth (*b*) histograms. The left and right *y*-axes show the density of small voids (≤3 μm) and larger voids (>3 μm), respectively. **d** Lifetime (*d-b*) histograms. The number of void pairs shown on the *y*-axis is normalized by the scanning volume.

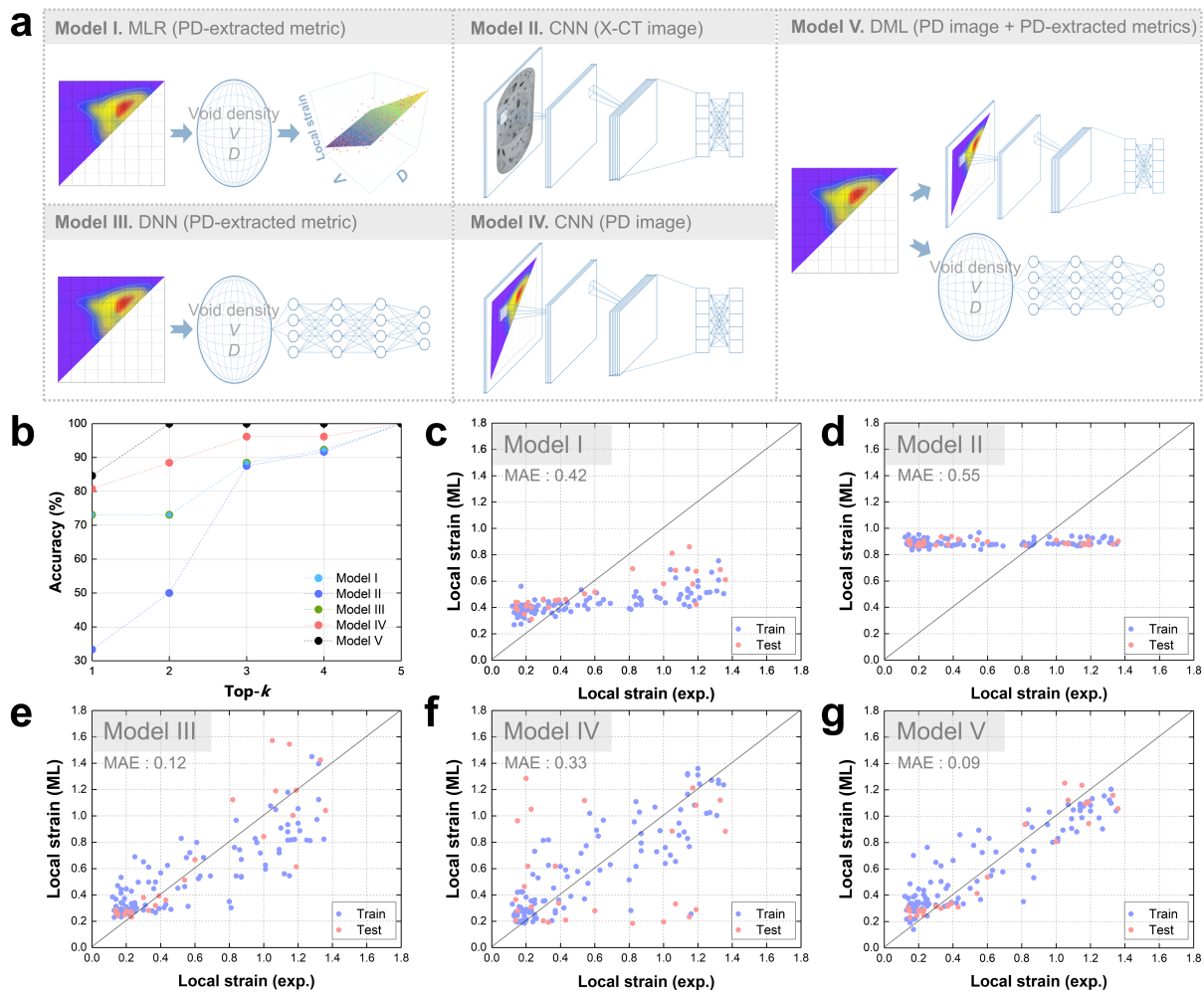

**Fig. 5**. **ML results using the tensile dataset. a** Schematics of the ML architecture and input types for Models **I**–**V**. **b** Top-*k* accuracies for the classifications of local strains. **c-g** Graphs comparing the local strains from experiments and ML predictions for Models **I**–**V**.

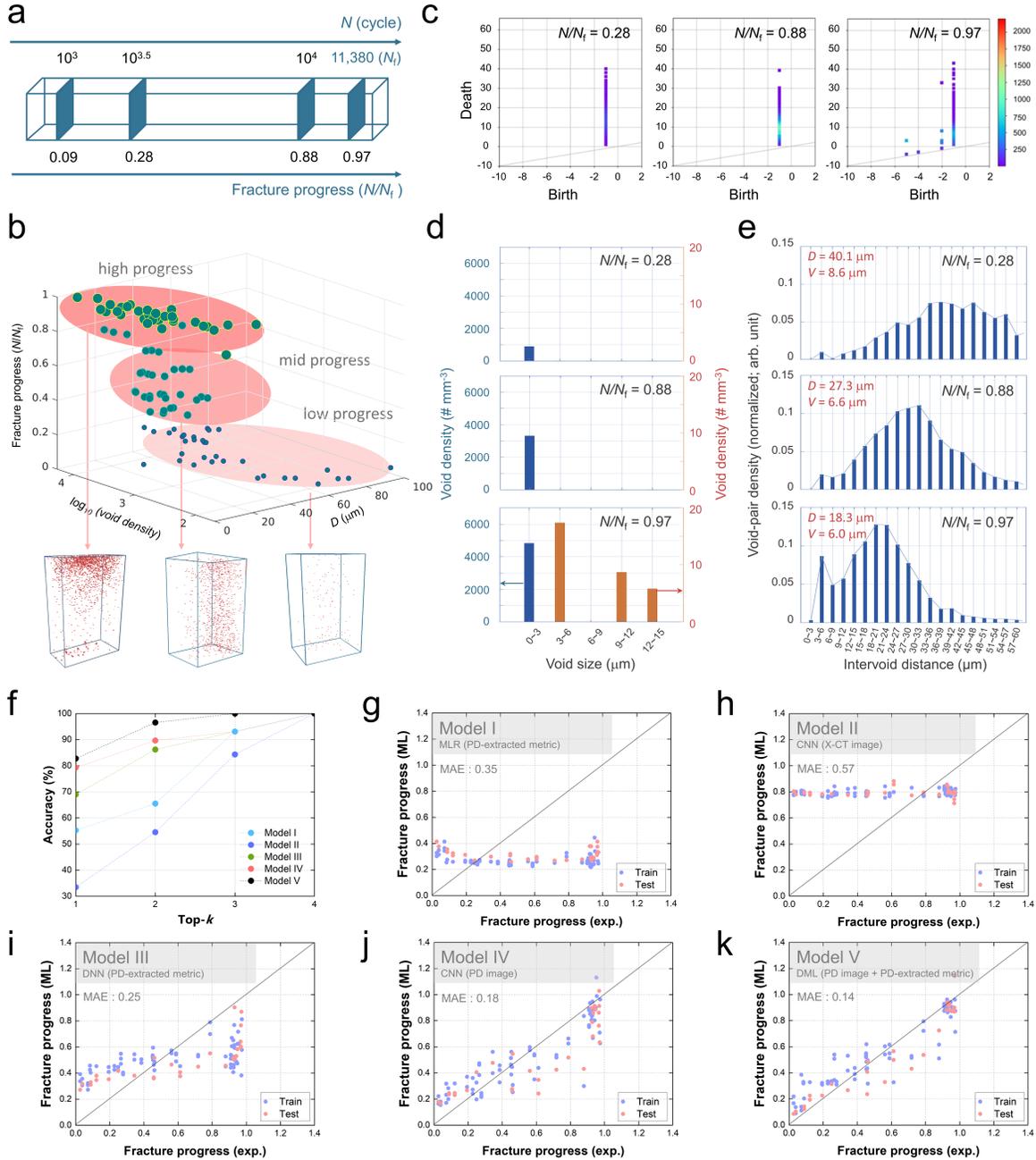

**Fig. 6. Application of the developed method to the fatigue fracture dataset. a** Scheme of data collection from fatigue tests. The X-CT data were obtained discretely at cycles of $10^{3.0}$, $10^{3.5}$, $10^{4.0}$, and $10^{4.5}$ and so forth until the specimen fractured. **b** The fatigue dataset with X-CT images. The graph shows the fracture progress ($N/N_f$ on the $z$-axis) as a function of the void density (# mm$^{-3}$, log scale) and $D$ (μm). **c** The $0^{th}$ PDs for representative samples with low, mid, and high progress. **d** Birth ($b$) histograms. **e** Lifetime ($d$-$b$) histograms. **f** Top-$k$ accuracies for the classification of the fracture progress. **g-k** Graphs comparing fracture progress from experiments and ML predictions for Models **I**–**V**.

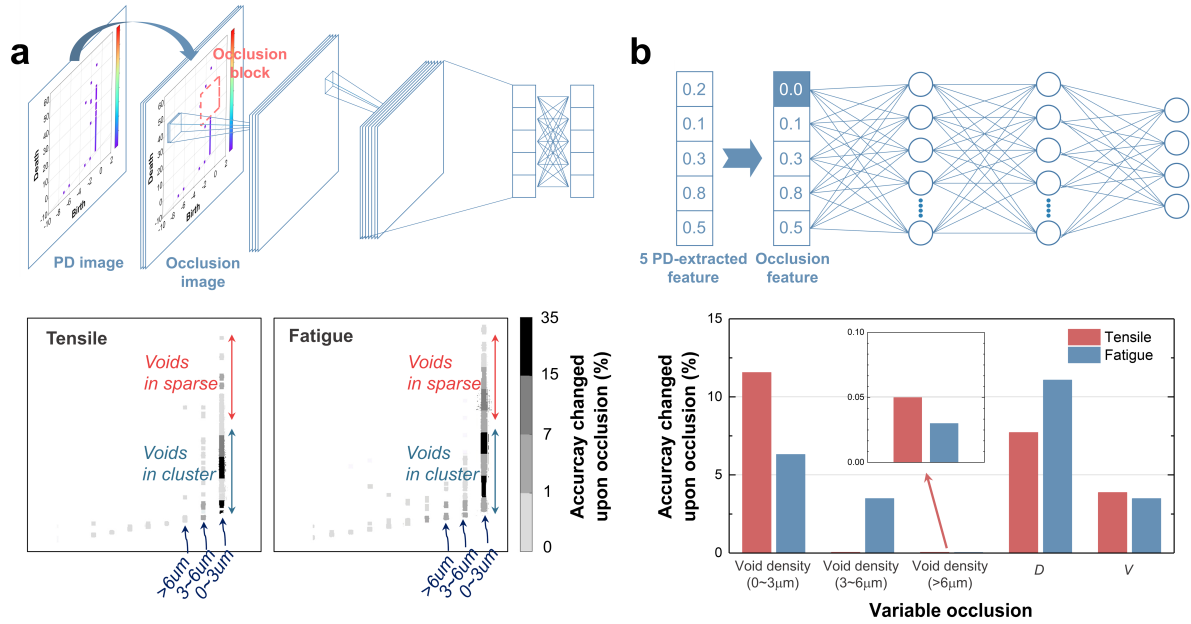

**Fig. 7**. **OSA results. a** Scheme illustrating the OSA process for Model **IV** (CNN based on PD images) and the corresponding accuracy changes with occlusions. **b** Scheme describing the OSA process for Model **III** (DNN based on PD-extracted metrics) and the corresponding accuracy changes with occlusions.

Supplementary information for

# Predicting failure characteristics of structural material via deep learning based on nondestructive void topology

*Leslie Ching Ow Tiong,[1][†] Gunjick Lee,[2][†] Seok Su Sohn,[2][*] and Donghun Kim[1][*]*

[1]Computational Science Research Center

Korea Institute of Science and Technology, Seoul 02792, Republic of Korea

[2]Department of Materials Science and Engineering

Korea University, Seoul 02841, Republic of Korea

†These authors contributed equally.

*Correspondence to: donghun@kist.re.kr (D.K.); sssohn@korea.ac.kr (S.S.S.)

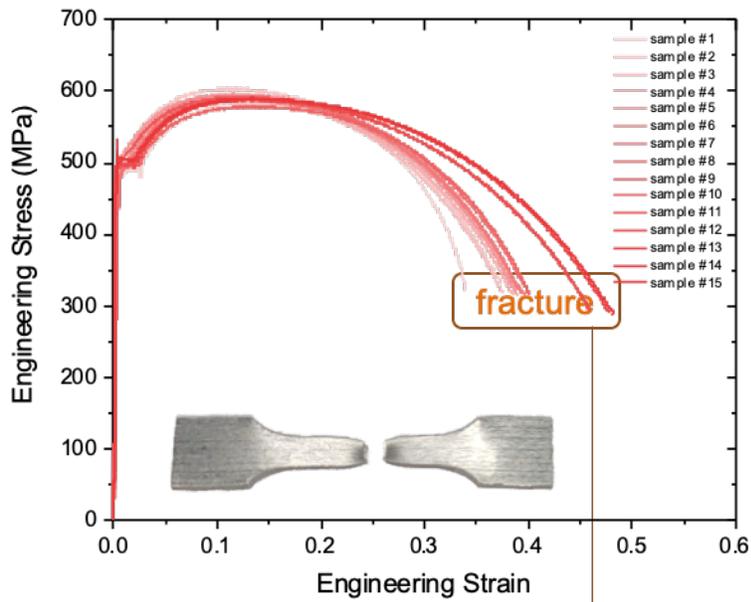

**Supplementary Fig 1. Engineering stress-strain curves of 15 tested ferritic steel samples.** The total elongation at the sample break is measured at 40.7±3.7%.

**Supplementary Table 1. Configurations of the CNN and DNN.** *f* refers to the feature map, *k* is defined as filter size, **and** C is defined as the number of classes.

| CNN | | DNN | |
|---|---|---|---|
| *Layers* | *Configurations* | *Layers* | *Configurations* |
| Input | 150×150×3 | Input | 1×5 |
| $Conv_1$ | *f*: 64@150×150, *k*: 2×2, *ReLu* | $FC_1$ | 1×64, *ReLu*, *Dropout* |
| Max-pooling | 2×2 | $FC_2$ | 1×256, *ReLu*, *Dropout* |
| $Conv_2$ | *f*: 128@75×75, *k*: 2×2, *ReLu* | $FC_2$ | 1×32, *ReLu*, *Dropout* |
| Max-pooling | 2×2 | Output | 1×C |
| $Conv_3$, $Conv_4$ | *f*: 256@38×38, *k*: 2×2, *ReLu* | | |
| Max-pooling | 2×2 | | |
| $Conv_5$, $Conv_6$ | *f*: 512@19×19, *k*: 2×2, *ReLu* | | |
| Max-pooling | 2×2 | | |
| $Conv_7$, $Conv_8$ | *f*: 512@9×9, *k*: 2×2, *ReLu* | | |
| Max-pooling | 2×2 | | |
| flatten | 512×5×5 | | |
| $FC_1$ | 1×4096, *ReLu*, *Dropout* | | |
| $FC_2$ | 1×4096, *ReLu*, *Dropout* | | |
| Output | 1×C | | |

**Supplementary Table 2. Configurations of the sub-networks in DML model.** *f* refers to the feature map, *k* is defined as filter size, **and** C is defined as the number of classes.

| Conv-based network in DML | |
|---|---|
| *Layers* | *Configurations* |
| Input | 150×150×3 |
| $Conv_1$ | *f*: 64@150×150, *k*: 3×3, *ReLu* |
| Max-pooling | 2×2 |
| $Conv_2$ | *f*: 128@75×75, *k*: 3×3, *ReLu* |
| Max-pooling | 2×2 |
| $Conv_3$, $Conv_4$ | *f*: 256@38×38, *k*: 3×3, *ReLu* |
| Max-pooling | 2×2 |
| $Conv_5$, $Conv_6$ | *f*: 512@19×19, *k*: 3×3, *ReLu* |
| Max-pooling | 2×2 |
| $Conv_7$, $Conv_8$ | *f*: 512@10×10, *k*: 3×3, *ReLu* |
| Max-pooling | 2×2 |
| flatten | 512×5×5 |
| $FC_1$ | 1×4096, *ReLu*, *Dropout* |
| $FC_2$ | 1×4096, *ReLu*, *Dropout* |
| Output | 1×C |

| Deep FC-based network in DML | |
|---|---|
| *Layers* | *Configurations* |
| Input | 1×5 |
| $FC_1$ | 1×64, *ReLu*, *Dropout* |
| $FC_2$ | 1×128, *ReLu*, *Dropout* |
| $FC_2$ | 1×32, *ReLu*, *Dropout* |
| Output | 1×C |

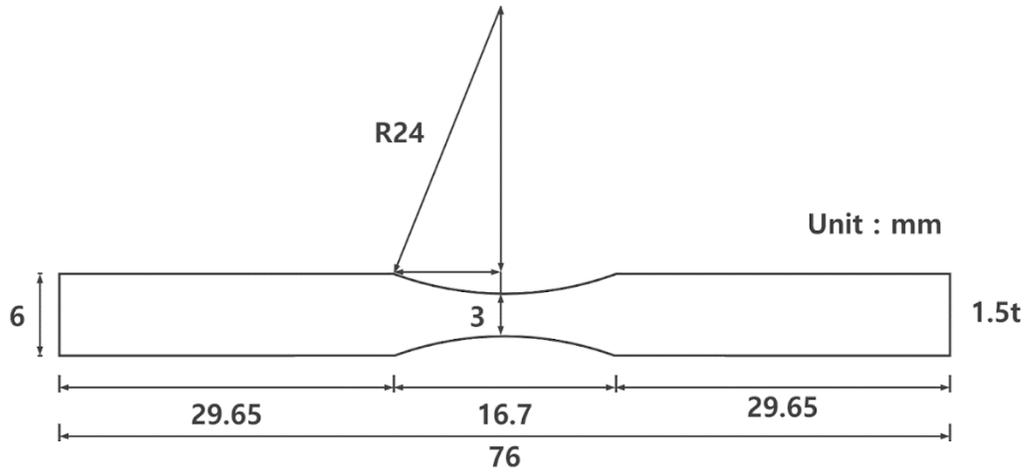

**Supplementary Fig 2. Detailed dimensions of fatigue testing specimen.**

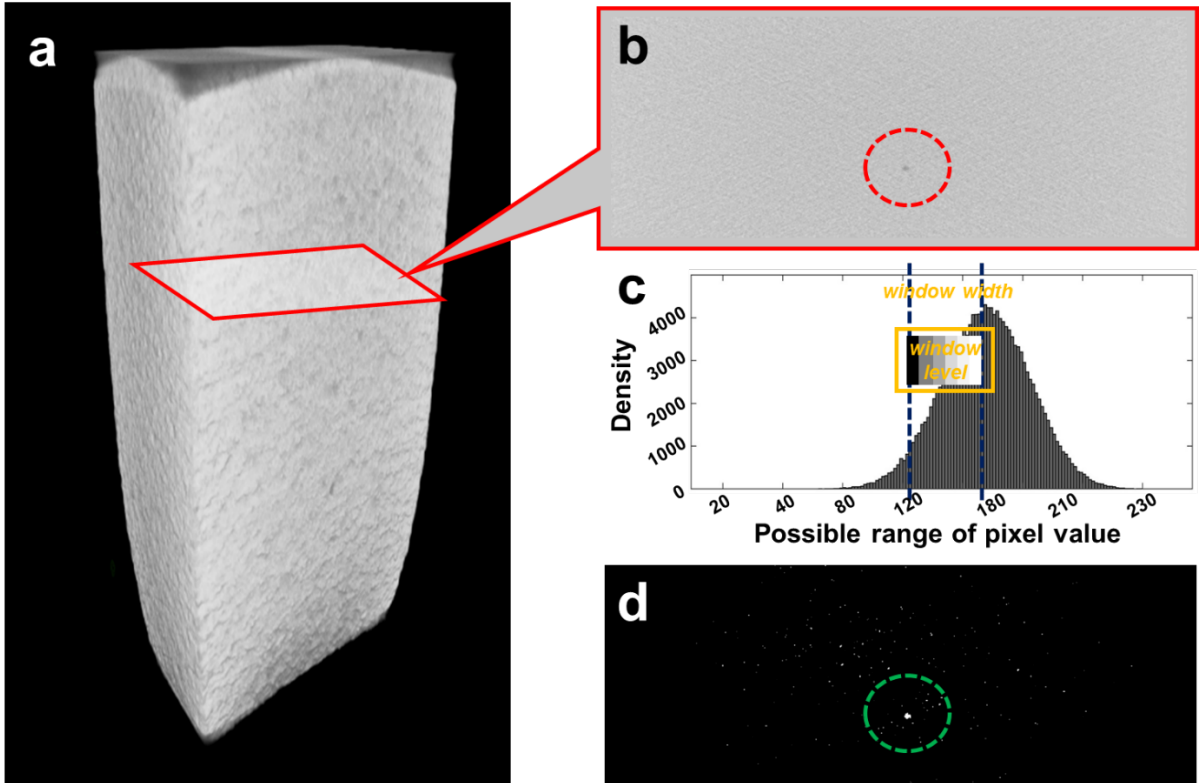

Supplementary Fig 3. Transformation of raw X-CT images into binary format images for extracting void appearances. a An exemplary raw 3D X-CT image. b A 2D slice view. c Demonstration of window-level processing. d Transformed binary format image.